\documentclass[10pt,tightenlines,eqsecnum,floats,aps,amsmath,amssymb,nofootinbib,prd,showpacs]{revtex4}

\usepackage{amssymb}
\usepackage{stmaryrd}
\usepackage{amsmath}
\usepackage{amsfonts}
\usepackage{mathrsfs}

\usepackage{amsmath,amssymb,amsfonts}
\usepackage{graphicx}

\def\be{\begin{equation}}
\def\ee{\end{equation}}
\def\ba{\begin{eqnarray}}
\def\ea{\end{eqnarray}}

\def\z{\zeta}
\def\Hk{\mathcal{H}_{\rm kin}} 
\def\Hkg{\mathcal{H}_{\rm kin}^{\rm grav}} 
\def\Hkm{\mathcal{H}_{\rm kin}^{\rm matt}} 
\def\a{\alpha}
\def\c{\sigma}

\def\z{\zeta}
\def\e{\epsilon}
\def\T{\widehat{\Theta}}
\def\TE{\widehat{\Theta}_{\rm E}}
\def\TL{\widehat{\Theta}_{\rm L}}
\def\D{\Delta}

\newcommand{\ket}[1]{\vert{#1} \rangle} 
\newcommand{\bra}[1]{\langle{#1}\,\vert} 
\newcommand{\obra}[1]{({#1}\vert} 
\newcommand{\inner}[2]{{\langle {#1} \vert  {#2} \rangle}} 
\newcommand{\opelem}[3]{{\langle {#1} \vert  {#2}  \vert  {#3} \rangle}} 
\newcommand{\norm}[1]{\|{#1}\|}
\newcommand{\com}[1]{\Psi_{z_{#1}}}
\newcommand{\cog}[1]{{\Psi}_{\zeta_{#1}}}
\newcommand{\cow}[1]{\Psi_{\eta_{#1}}}

\newcommand{\eqnref}[1]{Eq.~(\ref{#1})}

\begin{document}

\title{Coherent State Functional Integral in Loop Quantum Cosmology: Alternative Dynamics}

\author{Li Qin}\email{qinli051@163.com}
\author{Yongge Ma\footnote{Corresponding author}}\email{mayg@bnu.edu.cn}
\affiliation{Department of Physics, Beijing Normal University, Beijing 100875, China}

\begin{abstract}
Coherent state functional integral for the minisuperspace model of
loop quantum cosmology is studied. By the well-established canonical
theory, the transition amplitude in the path integral
representation of loop quantum cosmology with alternative
dynamics can be formulated through group averaging. The
effective action and Hamiltonian with higher-order quantum
corrections are thus obtained. It turns out that for a non-symmetric
Hamiltonian constraint operator, the \emph{Moyal (star)-product} emerges naturally in the
effective Hamiltonian. For the corresponding symmetric Hamiltonian operator, the resulted effective theory
implies a possible quantum cosmological effect in large scale limit
in the alternative dynamical scenario, which coincides with the result in canonical approach. Moreover, the first-order modified Friedmann equation still contains the particular information of alternative dynamics and hence admits the possible phenomenological distinction between the different proposals of quantum dynamics.\\

{\textbf{Keywords}: Loop quantum cosmology; Coherent state functional integral; Effective theory}
\pacs{98.80.Qc, 04.60.Pp, 03.65.Sq, 11.10.Nx}

\end{abstract}

\maketitle

\section{Introduction}
In the last two decades, considerable progress has been made in loop quantum gravity (LQG),
which is a background independent approach to quantum gravity \cite{Rovelli,A-L,Thiemann,Han}.
The starting point of LQG is the Hamiltonian \emph{connection dynamics} of GR
rather than the ADM formalism. By taking the holonomy of
$su(2)$-connection $A_{a}^{i}$ and flux of densitized triad
$E^{b}_{j}$ as basic variables, the quantum kinematical framework of
LQG has been rigorously constructed, and the Hamiltonian constraint
operator can also be well defined to represent quantum dynamics.
Moreover, a few physically significant results have been obtained in
the minisuperspace models of loop quantum cosmology (LQC)
\cite{Boj,Ash-view}. The most interesting one is the
\emph{resolution of big bang singularity} in LQC \cite{Boj2,aps1,aps2,aps3}.
Besides the canonical formalism, the so-called spin foam models were
proposed as the path integral formalism of LQG \cite{Rovelli}.
However, whether the two approaches are equivalent to each other is
a longstanding open question. Thanks to the development of LQC, we
have a much simpler theory to address this question. As
symmetry-reduced models, there are only finite number of degrees of
freedom in LQC. Following the conventional method in quantum
mechanics, one can find the path integral formalism of LQC starting
with the canonical formalism. This approach has been implemented by
a series of works \cite{Henderson-1,Henderson-2,Henderson-3,Huang} with the scheme of
simplified LQC \cite{robust}. Here one employed the complete basis
of eigen-states of the volume operator to formulate a path integral
with somehow \emph{descrete-steps}, which inherited certain
properties of spin foams \cite{RV2}. Moreover, the first-order
effective action for the path integral was derived by this approach
\cite{Henderson-3,Huang}, which implied the origin of singularity
resolution of LQC in the path integral representation. In canonical
LQC, the effective Hamiltonian constraint with higher-order quantum
corrections could even be obtained by the semiclassical analysis
using coherent states, which implied a possible effect of quantum
gravity on large scale cosmology \cite{Ding,Yang,ma}. It is thus
interesting to see whether the effective Hamiltonian can be
confirmed by some path integral representation. Since the
higher-order corrections of the Hamiltonian come from the quantum
fluctuations, a natural attempt to achieve them is to employ
coherent state path integral \cite{Brown-book}.

In LQC, the Hamiltonian constraint equation is usually presented to Klein-Gordon like equation by coupling with a massless scalar field, where the corresponding gravitational Hamiltonian operator, as some multiplication of several self-adjoint operators, is non-symmetric in the kinematical Hilbert space \cite{aps2,aps3,robust}. While this treatment is essential
in order to obtain the physical states satisfying the constraint
equation, it also provides elegant physcial models to examine the
so-called \emph{Moyal} $*$\emph{-product} in quantum mechanics. At
the very beginning, Moyal proposed the $*$\emph{-product} in order
to clarify the role of statistical concepts in quantum mechanics
system \cite{Moyal}. Then this idea were generalized to many
situations including quantum spacetime itself. In canonical quantum
theories, the $*$\emph{-product} can also be understood by coherent
state approach \cite{book-fuzzy}. Thus it is also possible and
desirable to derive the $*$\emph{-product} in coherent state
functional integral approach. This idea has been accomplished in the
WDW quantum cosmology and LQC \cite{Qin}. In these
models, the \emph{Euclidean} and \emph{Lorentz} terms in the gravitational Hamiltonian
constraint are combined together  since they are proportional to
each other in spatially flat and homogeneous cases. However, this is
impossible in the full theory, where the Lorentz term has to be
quantized in a form quite different from the Euclidean one \cite{Thiemann,Han}. This
kind of quantization procedure which kept the distinction of the two
terms was proposed as alternative dynamics for LQC \cite{Yang}. Hence we
will study the coherent state functional integral in spatially flat
isotropic FRW cosmology coupled with a massless scalar field $\phi$
in the alternative quantization framework.

\section{Coherent functional integrals}
We consider the following Hilbert-Einstein action of gravity coupled with a massless scalar field:
 \be\label{eqn:action}
 S=\frac{1}{16\pi{G}}\int{d^4x}\sqrt{-g}R
 +\frac{1}{2}\int{d^4x}\sqrt{-g}g^{\mu\nu}\phi_{,\nu}\phi_{,\nu}.
 \ee
In the spatially flat model of FRW cosmology, we fix a space-like sub-manifold $S$,
which is topologically $\mathbb{R}^3$ and equipped with Cartesian
coordinates $x^i(i=1,2,3)$ and a fiducial flat metric
${}^o\!q_{ab}$. The physical 3-metric $q_{ab}$ is then determined by
a scale factor $a$ satisfying $q_{ab}=a^2{}^o\!q_{ab}$. It is
convenient to introduce an elementary cell $\mathcal{V}$ and
restrict all integrations to this cell in Hamiltonian analysis. The
volume of $\mathcal{V}$ with respect to ${}^o\!q_{ab}$ is denoted as
$V_o$. As in the full loop quantum gravity, we employ the new
canonical variables $(A_a^i,E_i^a)$ \cite{mathematical structure}.
Due to the homogeneity and isotropy, we can fix a set of orthonormal
cotriad and triad $({}^o\!\omega_a^i,{}^o\!e^a_i)$ compatible with
${}^oq_{ab}$ and adapted to $\mathcal{V}$. Then the basic canonical
variables take the simple form
$A_a^i=cV_o^{-(1/3)}{}^o\!\omega_a^i$,
$E_i^a=p\sqrt{{}^o\!q}V_o^{-(2/3)}{}^o\!e^a_i$ and thus are reduced
to $(c,p)$ with the Poisson bracket: $\{c,p\}=8\pi G\gamma/3$, where
$\gamma$ is the Barbero-Immirzi parameter. Following the
\emph{$\bar{\mu}$-schem} of \emph``improved dynamics" \cite{aps3},
the regulator $\bar{\mu}$ used in holonomies is given by
$\bar{\mu}=\sqrt{{\D}/{|p|}}$, where
$\D=4\sqrt{3}\pi\gamma{\ell}_{\textrm{p}}^2$ is a minimum nonzero
eigenvalue of the area operator \cite{Ash-view}. In order to do the
semiclassical analysis, it is convenient to introduce new
dimensionless conjugate variables \cite{robust,Yang}:
 \be\label{eqn:b and v}
 b:=\frac{\bar{\mu}c}{2}
 ,\quad
 v:=\frac{\text{sgn}(p)|p|^{\frac{3}{2}}}{2\pi\gamma{\ell}^2_{\textrm{p}}\sqrt{\D}},
 \ee
with the Poisson bracket $\{v,b\}=-\frac{1}{\hbar}$ , where the
Planck length $\ell_{\textrm{p}}$ is given by
$\ell_{\textrm{p}}^2=G\hbar$. From the matter part of action
(\ref{eqn:action}), we can get
$p_{\phi}=\frac{a^3V_o\dot{\phi}^2}{2}$ and the poisson bracket:
$\{\phi,p_{\phi}\}=1$. The kinematical Hilbert space of the quantum
theory is supposed to be a tensor product of the gravitational and
matter parts. In LQC, one employed the standard
\emph{Schr\"{o}dinger} representation for matter to construct
Hilbert space $\Hkm$, while gravity was quantized by the polymer-like
representation \cite{mathematical structure}. Thus quantum states in
the gravitational Hilbert space of LQC are functions ${\Psi}(v)$ of
$v$ with support on a countable number of points and with finite
norm $\norm{{\Psi}}^2:=\sum_{v}|{\Psi}(v)|^2$ \cite{shadow}. Hence
the inner product is defined by a \emph{Kronecker delta}
$\inner{v'}{v}=\delta_{v',v}$. The basic operators act on a quantum
state ${\Psi}(v,\phi)$ in the kinematical Hilbert space $\Hkg$ as:
 \be\label{eqn:v,b action}
 \hat{v}{\Psi}(v,\phi)=v{\Psi}(v,\phi),\quad
 \widehat{e^{ib}}{\Psi}(v,\phi)={\Psi}(v+1,\phi).
 \ee
To obtain the physical states, one has to solve the quantum
Hamiltonian constraint equation:
 \be\label{eqn:quantum constrain LQC}
 -\hat{C}\cdot{\Psi}(v,\phi)=\left(-\frac{\hat{p}_{\phi}^2}{\hbar^2}+\T\right){\Psi}(v,\phi)=0,
 \ee
where $\T\equiv\TE+\TL$ is a positive second-order difference
operator defined by:
\begin{subequations}
 \ba
 \TE\cdot\Psi(v,\phi)&=&\frac{3\pi{G}\gamma^2}{4}
 \left[v(v+2){\Psi}(v+4,\phi)-2v^2{\Psi}(v,\phi)+v(v-2){\Psi}(v-4,\phi)\right]\label{eqn:action
 TE},\\
 \TL\cdot\Psi(v,\phi)&=&-\frac{3\pi{G}(1+\gamma^2)}{16}
 \left[v(v+4){\Psi}(v+8,\phi)-2v^2{\Psi}(v,\phi)+v(v-4){\Psi}(v-8,\phi)\right]\label{eqn:action
 TL}.
  \ea
\end{subequations}
Here we use the \emph{alternative quantization} scheme proposed in
\cite{Yang} in which the \emph{Euclidean} and \emph{Lorentz} terms
in the gravitational \emph{Hamiltonian constraint} are treated separately. Together
with the simplified treatment in \cite{robust}, we can get the
operators in Eqs. (\ref{eqn:action  TE}) (\ref{eqn:action  TL})
corresponding to the \emph{Euclidean} and \emph{Lorentz} terms
respectively. Solutions to the constraint equations and their
physical inner products can be obtained through the group averaging
procedure. Now we concern about \emph{coherent state functional
integrals}. The (\emph{generalized}) coherent state of the matter
part is labeled by a complex variable
$z_o:=\frac{1}{\sqrt{2}\c}(\phi_o+\frac{i}{\hbar}\c^2p_{\phi_o})$
and defined by
 \be\label{eqn:coherent state matter}
 \ket{\com{o}}:=\int_{-\infty}^{\infty} d\phi\
 e^{-\frac{(\phi-\phi_o)^2}{2\c^2}}e^{\frac{i}{\hbar}p_{\phi_o}(\phi-\phi_o)}\ket{\phi},
 \ee
which is the eigenstate of the \emph{annihilation} operator
$\hat{z}=\frac{1}{\sqrt{2}\c}(\hat{\phi}+\frac{i}{\hbar}\,\hat{p}_{\phi}\c^2)$,
where $\c$ describes the width of the wave-packet or quantum
fluctuation. It satisfies the key properties of a coherent state,
namely, saturation of Heisenberg's uncertainty relation, resolution
of identity and peakness property.  On the other hand, due to the
\emph{polymer-like} structure, the coherent state of LQC is
different from that of the matter part. Here one can define
$\z_o=\frac{1}{\sqrt2d}(v_o+ib_od^2)$ to label the
\emph{generalized} coherent state \cite{shadow,Ding}:
 \be\label{eqn:coherent state gravity}
 \obra{\cog{o}}:=\sum_{v\in\mathbb{R}}
 e^{-\frac{(v-v_o)^2}{2d^2}}e^{-ib_o(v-v_o)}\obra{v},
 \ee
where $d$ is the characteristic \emph{width} of the wave packet and
$1\ll{d}\ll{v_o}$ because of the semiclassical feature. For
practical use, one defines the projection of this state on some
lattice of variable $v$, saying the \emph{shadow state}
\cite{shadow}:
 \be\label{eqn:shadow}
 \ket{\cog{o}}^{\rm shad}:=\sum_{k=-\infty}^{\infty}
 e^{-\frac{(k-v_o)^2}{2d^2}}e^{ib_o(k-v_o)}\ket{k},\quad
 k\in\mathbb{Z},
 \ee
where we chose the regular lattice $\{v=k,k\in\mathbb{Z}\}$. This
shadow state also has the analogous properties of a coherent state. The resolution of identity now reads
 \be\label{eqn:gravity identity of coherent}
 \int_{-\infty}^{\infty}dv_o\int_{-\pi}^{\pi}\frac{db_o}{2\pi}
 \frac{\ket{\cog{o}}\bra{\cog{o}}}{\inner{\cog{o}}{\cog{o}}}
 =\sum_{k=-\infty}^{\infty}\ket{k}\bra{k}\equiv\mathbb{I},
 \ee
where the identity $\mathbb{I}$ is in the subspace in which the
states have support only on the regular lattice. The whole coherent
state of LQC reads $\ket{\com{o}}\ket{\cog{o}}\equiv
\ket{\com{o}}\otimes\ket{\cog{o}}$.

In the path integral of the conventional non-relativistic quantum
mechanics, one needs to compute the matrix element of the
\emph{evolution operator} $e^{-i\Delta{t}\hat{H}}$ within the time
interval $\Delta{t}$. However, the situation of cosmology in GR is
very different, since we are considering totally constrained systems and the
operator $\hat{C}$ is not a \emph{true Hamiltonian}. Instead, we
start from the physical inner product, i.e., the transition
amplitude, of coherent states with \emph{normalization}:
 \be\label{eqn:physical inner coherent}
  A([\Psi_{f}],[\Psi_{i}])\equiv\frac{\bra{\cow{f}}\bra{\com{f}}
   \int_{-\infty}^{\infty} d\a\ e^{i\a\hat{C}}\ket{\com{i}}\ket{\cow{i}}}
 {\norm{\Psi_{\eta_f}}\norm{\Psi_{z_f}}\norm{\Psi_{z_i}}\norm{\Psi_{\eta_i}}}.
  \ee
To calculate the \emph{transition amplitude}, we split a fictitious
\emph{time interval} $\Delta{\tau}=1$ into $N$ pieces
$\e=\frac{1}{N}$. To deal with the parameter $\a$ in group averaging
procedure, we employ the trick in \cite{Huang} to generalize the one
single group averaging to multiple ones, i.e.,
 \ba
 &&\lim_{\a_o\rightarrow\infty}\int_{-\a_o}^{\a_o}d\a~e^{i\a\hat{C}}\ket{\Psi_{\rm{kin}}}\\
 &=&\lim_{\tilde{\a}_{N_o},\cdots,\tilde{\a}_{1_o}\rightarrow\infty}
 \frac{1}{2\tilde{\a}_{N_o}}\int_{-\tilde{\a}_{N_o}}^{\tilde{\a}_{N_o}}d\tilde{\a}_N\cdots
 \frac{1}{2\tilde{\a}_{2_o}}\int_{-\tilde{\a}_{2_o}}^{\tilde{\a}_{2_o}}d\tilde{\a}_{2}
 \int_{-\tilde{\a}_{1_o}}^{\tilde{\a}_{1_o}}d\tilde{\a}_{1}
 e^{i(\tilde{\a}_{N}+\cdots+\tilde{\a}_{1})\hat{C}}\ket{\Psi_{\rm{kin}}},~
 \forall~\ket{\Psi_{\rm{kin}}}\in\Hk.\label{eqn:group average N}
 \ea
In order to trace the power for expansion, we re-scale the
parameters by $\tilde{\a}_n=\e\a_n(n=1,\cdots,N)$ and thus rewrite
the exponential operator as:
$e^{i\sum_{n=1}^{N}\e\a_n\hat{C}}=\prod_{n=1}^{N}e^{i\e\a_n\hat{C}}$.
Inserting $N$ times of coherent states resolution of identity of
$\ket{\com{o}}$ and \eqnref{eqn:gravity identity of coherent},
\eqnref{eqn:physical inner coherent} can be casted into
 \be\label{eqn:amplitude2}
 A([\Psi_{f}],[\Psi_{i}])
 =\lim_{\a_{N_o},\cdots,\a_{1_o}\rightarrow\infty}
 \frac{1}{2{\a}_{N_o}}\int_{-{\a}_{N_o}}^{{\a}_{N_o}}d{\a}_N\cdots
 \frac{1}{2{\a}_{2_o}}\int_{-{\a}_{2_o}}^{{\a}_{2_o}}d{\a}_{2}
 \cdot\e\int_{-{\a}_{1_o}}^{{\a}_{1_o}}d{\a}_{1}~
 A^{\textrm{matt}}_{N}~A^{\textrm{grav}}_{N},
\ee
where
  \begin{subequations}
 \ba
 A^{\rm matt}_{N}=\int_{-\infty}^{\infty}d\phi_{N-1}\dots d\phi_{1}
 \int_{-\infty}^{\infty} \frac{dp_{\phi_{N-1}}}{2\pi\hbar}\dots\frac{dp_{\phi_1}}{2\pi\hbar}
 \prod_{n=1}^{N}
 \frac{\opelem{\com{n}}{e^{i\e\a_n\frac{\hat{p}_{\phi}^2}{\hbar^2}}}{\com{n-1}}}{\norm{\com{n}}\norm{\com{n-1}}},
 \label{eqn:A matter}\\
 A^{\rm grav}_{N}=\int_{-\infty}^{\infty}dv_{N-1}\dots dv_{1}
 \int_{-\pi}^{\pi} \frac{db_{N-1}}{2\pi}\dots\frac{db_1}{2\pi}
 \prod_{n=1}^{N}\frac{\opelem{\cow{n}}{e^{-i\e\a_n\hat{\Theta}}}{\cow{n-1}}}{\norm{\cow{n}}\norm{\cow{n-1}}},
 \label{eqn:A gravity}
 \ea
 \end{subequations}
with $z_N\equiv z_f,z_0\equiv z_i,\eta_N\equiv\eta_f, \textrm{and}
~\eta_0\equiv\eta_i$. Notice that the characteristic widths $\c$ and
$d$ at different steps are not necessarily the same. So we have to
denote $\c_n$ and $d_n$ in the semiclassical states $\ket{\com{n}}$
and $\ket{\cog{n}}$ respectively at the ``n-step". Now the main task
is to calculate the matrix elements of the exponential operators on
coherent states. The exponential operator $e^{i\e\a_n\hat{C}}$ can
be expanded as $1+i\e\a_n\hat{C}+\mathcal{O}(\e^2)$. For the purpose
of a concise writing, we introduce some intermediate-step notations:
 \be\nonumber
   \overline{p}_{\phi_n} \equiv
  \frac{\c^2_np_{\phi_n}+\c^2_{n-1}p_{\phi_{n-1}}}{\c^2_n+\c^2_{n-1}},\quad
  \overline{\c_n^2} \equiv
  \frac{2\c^2_n\c^2_{n-1}}{\c^2_n+\c^2_{n-1}}.
 \ee
The product of the matrix elements in \eqnref{eqn:A matter} can be calculated as \cite{Qin}:
 \ba
 \prod_{n=1}^{N}\frac{\opelem{\com{n}}{e^{i\e\a_n\frac{\hat{p}_{\phi}^2}{\hbar^2}}}{\com{n-1}}}
{\norm{\com{n}}\norm{\com{n-1}}}
 =\left(\prod_{n=1}^{N}
 \frac{\inner{\com{n}}{\com{n-1}}}{\norm{\com{n}}\norm{\com{n-1}}}\right)
 \exp\Big[\frac{i\e\a_n}{\hbar^2}
 \sum_{n=1}^{N}\Big(p_{\phi_{n-1}}^2+\frac{\hbar^2}{\c_n^2+\c_{n-1}^2}\Big)\Big],\label{eqn:A-N matter}
 \ea
where the product of series $\inner{\com{n}}{\com{n-1}}$ can be
expressed as
 \ba
 &&\prod_{n=1}^{N}
 \frac{\inner{\com{n}}{\com{n-1}}}{\norm{\com{n}}\norm{\com{n-1}}}
 =\exp\left[\frac{\phi_N^2+p_{\phi_N}^2\c_{N+1}^2\c_N^2/\hbar^2}{2(\c_{N+1}^2+\c_N^2)}
 -\frac{\phi_0^2+p_{\phi_0}^2\c_{1}^2\c_0^2/\hbar^2}{2(\c_{1}^2+\c_0^2)}\right]
 \left(\prod_{n=1}^{N}\sqrt{\frac{2\c_n\c_{n-1}}{\c_n^2+\c_{n-1}^2}}\right)\nonumber\\
 &&\cdot\exp\Big[\e\sum_{n=1}^{N}\Big(-\frac{2(\c_{n+1}^2+\c_n^2)\phi_n\frac{\phi_n-\phi_{n-1}}{\e}
 -(\c_{n+1}+\c_{n-1})\frac{\c_{n+1}-\c_{n-1}}{\e}\phi_n^2}
 {2(\c_{n+1}^2+\c_{n}^2)(\c_n^2+\c_{n-1}^2)}
 +\frac{i}{\hbar}\overline{p}_{\phi_n}
 \frac{\phi_n-\phi_{n-1}}{\e}\nonumber\\
 &&\quad\quad\quad\quad\quad\quad\quad
 -\frac{1}{4\hbar^2}\frac{4(\c_{n+1}^2\c_n^2\c_{n-1}^2+\c_n^4\c_{n-1}^2)p_{\phi_n}
 \frac{p_{\phi_n}-p_{\phi_{n-1}}}{\e}
 +2\c_{n}^4(\c_{n+1}+\c_{n-1})\frac{\c_{n+1}-\c_{n-1}}{\e}p_{\phi_n}^2}
 {(\c_{n+1}^2+\c_n^2)(\c_n^2+\c_{n-1}^2)}\Big)\Big].\label{eqn:prod
 matter}
 \ea
Here we introduced a \emph{virtual width} $\c_{N+1}$ by hand,
satisfying $\c_{N+1}-\c_{N}=\c_{N}-\c_{N-1}$, in order to get the
tidy sum in the exponential position. In the limit of
$N\rightarrow\infty$, $\c_{N+1}$ will approach $\c_N\equiv\c_f$ and
hence does not effect the quantum dynamics.

For the gravitational part, careful calculations outlined in the Appendix yield
 \ba
 &&\prod_{n=1}^{N}\frac{\opelem{\cog{n}}{e^{-i\e\a_n\T}}{\cog{n-1}}}
{\norm{\cog{n}}\norm{\cog{n-1}}}
 =\left(\prod_{n=1}^{N}
 \frac{\inner{\cog{n}}{\cog{n-1}}}{\norm{\cog{n}}\norm{\cog{n-1}}}\right)
 \exp\Big[i\e\a_n\cdot3\pi{G}\nonumber\\
 &&\times\sum_{n=1}^{N}\Big(\gamma^2\Big(\big(\overline{v}^2_n+\frac{\overline{d^2_n}}{2}\big)
 \big(\sin^2{(2\overline{b}_n)}\big(1-\frac{8}{d^2_n+d^2_{n+1}}\big)+\frac{4}{d^2_n+d^2_{n-1}})
 +i\overline{v}_n\sin{(4\overline{b}_n)}\frac{2d^2_n}{d^2_n+d^2_{n-1}}\big(1-\frac{8}{d^2_n+d^2_{n+1}}\big)
 \Big)\nonumber\\
 &&\quad\quad-\frac{1+\gamma^2}{4}\Big(\big(\overline{v}^2_n+\frac{\overline{d^2_n}}{2}\big)
 \big(\sin^2{(4\overline{b}_n)}\big(1-\frac{32}{d^2_n+d^2_{n+1}}\big)+\frac{16}{d^2_n+d^2_{n-1}})
 +2i\overline{v}_n\sin{(8\overline{b}_n)}\frac{2d^2_n}{d^2_n+d^2_{n-1}}\big(1-\frac{32}{d^2_n+d^2_{n+1}}\big)
 \Big)\Big)\Big]\nonumber\\\label{eqn:prod LQC result}
 \ea
where
$\overline{v}_n\equiv\frac{d^2_{n-1}v_n+d^2_{n}v_{n-1}}{d^2_n+d^2_{n-1}},
 \overline{b}_n\equiv\frac{d^2_nb_n+d^2_{n-1}b_{n-1}}{d^2_n+d^2_{n-1}},
 \overline{d^2_n}\equiv\frac{2d^2_nd^2_{n-1}}{d^2_n+d^2_{n-1}}$, and
\ba
 &&\prod_{n=1}^{N}
 \frac{\inner{\cog{n}}{\cog{n-1}}}{\norm{\cog{n}}\norm{\cog{n-1}}}
 =\exp\left[\frac{v_N^2+b_{N}^2d_{N+1}^2d_N^2}{2(d_{N+1}^2+d_N^2)}
 -\frac{v_0^2+b_{0}^2d_{1}^2d_0^2}{2(d_{1}^2+d_0^2)}\right]
 \left(\prod_{n=1}^{N}\sqrt{\frac{2d_nd_{n-1}}{d_n^2+d_{n-1}^2}}\right)\nonumber\\
 &&\cdot\exp\Big[\e\sum_{n=1}^{N}\Big(-\frac{2(d_{n+1}^2+d_n^2)v_n\frac{v_n-v_{n-1}}{\e}
 -(d_{n+1}+d_{n-1})\frac{d_{n+1}-d_{n-1}}{\e}v_n^2}
 {2(d_{n+1}^2+d_{n}^2)(d_n^2+d_{n-1}^2)}
 +{i}\overline{b}_{n}
 \frac{v_n-v_{n-1}}{\e}\nonumber\\
 &&\quad\quad\quad\quad\quad\quad\quad\quad
 -\frac{4(d_{n+1}^2d_n^2d_{n-1}^2+d_n^4d_{n-1}^2)b_{n}\frac{b_{n}-b_{{n-1}}}{\e}
 +2d_{n}^4(d_{n+1}+d_{n-1})\frac{d_{n+1}-d_{n-1}}{\e}b_{_n}^2}
 {4(d_{n+1}^2+d_n^2)(d_n^2+d_{n-1}^2)}\Big)\Big].\label{eqn:prod LQC}
 \ea
Now we take the limit $N\rightarrow\infty$
and substitute $\int_{0}^{1}d\tau$ for $\sum_{n=1}^{N}\e$ to get the
functional integral formalism of the amplitude:
 \be\label{eqn:amplitude result WDW}
 A([\Psi_f][\Psi_i])=e^{\frac{1}{2}\left(|z_f|^2-|z_i|^2+|\z_f|^2-|\z_i|^2\right)}
 \int{\mathcal{D}\a}\int[\mathcal{D}\phi(\tau)][\mathcal{D}p_{\phi}(\tau)][\mathcal{D}v(\tau)][\mathcal{D}b(\tau)]
 e^{i(S_{\a}^{\rm{matt}}+S_{\a}^{\rm{grav}})},
 \ee
where \be\label{eqn:effective action matter}
 S_{\a}^{\rm{matt}}=\int_{0}^{1}d\tau\left(i\frac{d}{d\tau}\left(\frac{\phi^2}{4\c^2}\right)
 +i\frac{d}{d\tau}\left(\frac{\c^2p_{\phi}^2}{4\hbar^2}\right)+\frac{p_{\phi}\dot{\phi}}{\hbar}
 +\frac{\a}{\hbar^2}\left(p_{\phi}^2+\frac{\hbar^2}{2\c^2}\right)\right),
 \ee
\ba
 &&S_{\a}^{\rm{grav}}=\int_{0}^{1}d\tau\Big(i\frac{d}{d\tau}\left(\frac{v^2}{4d^2}\right)
 +i\frac{d}{d\tau}\left(\frac{d^2b^2}{4}\right)+{b\dot{v}}\nonumber\\
 &&\quad\quad\quad\quad\quad\quad\quad\quad
 +\a3\pi{G}\Big[\gamma^2\Big(v^2+\frac{d^2}{2}\Big)\Big(\Big(\sin^2{(2b)}\big(1-\frac{4}{d^2}\big)
 +\frac{2}{d^2}\Big)+iv\sin{(4b)}\big(1-\frac{4}{d^2}\big)\Big)\nonumber\\
 &&\quad\quad\quad\quad\quad\quad\quad\quad\quad\quad\quad\quad
 -\frac{1+\gamma^2}{4}
 \Big(v^2+\frac{d^2}{2}\Big)\Big(\Big(\sin^2{(4b)}\big(1-\frac{16}{d^2}\big)
 +\frac{8}{d^2}\Big)+2iv\sin{(8b)}\big(1-\frac{16}{d^2}\big)\Big)\Big]\Big).
 \label{eqn:effective action LQC}
 \ea
Here the ``dots" over $\phi$ and $v$ stand for the \emph{time
derivative} with respect to the \emph{fictitious time} $\tau$. The
\emph{functional measures} are defined on \emph{continuous paths} by
taking the limit of $N\rightarrow\infty$:
\begin{subequations}
\ba
 &&\int\mathcal{D}\a:=\lim_{N\rightarrow\infty}~~\lim_{\a_{N_o},\cdots,\a_{1_o}\rightarrow\infty}
 \frac{1}{2{\a}_{N_o}}\int_{-{\a}_{N_o}}^{{\a}_{N_o}}d{\a}_N\cdots
 \frac{1}{2{\a}_{2_o}}\int_{-{\a}_{2_o}}^{{\a}_{2_o}}d{\a}_{2}
 ~\frac{1}{N}\int_{-{\a}_{1_o}}^{{\a}_{1_o}}d{\a}_{1},\label{eqn:measure alpha}\\
 &&\int[\mathcal{D}\phi(\tau)][\mathcal{D}p(\tau)]:=\lim_{N\rightarrow\infty}
 \left(\prod_{n=1}^{N}\sqrt{\frac{2\c_n\c_{n-1}}{\c_n^2+\c_{n-1}^2}}\right)
 \int\prod_{n=1}^{N-1}\frac{d{\phi_n}dp_{\phi_n}}{2\pi\hbar}\ ,\label{eqn:measure-matter}\\
                 &&\int[\mathcal{D}v(\tau)][\mathcal{D}b(\tau)]:=\lim_{N\rightarrow\infty}
 \left(\prod_{n=1}^{N}\sqrt{\frac{2d_nd_{n-1}}{d_n^2+d_{n-1}^2}}\right)
 \int\prod_{n=1}^{N-1}\frac{d{v_n}db_n}{2\pi}\ .\label{eqn:measure-grav}
 \ea
  \end{subequations}
Ignoring the total derivatives with respect to $\tau$ in Eqs.
(\ref{eqn:effective action matter}) and (\ref{eqn:effective action
LQC}), we can read out the total effective Hamiltonian constraint
as:
 \ba
 {\mathscr{H}}_{\rm{eff}}=-\frac{p_{\phi}^2}{\hbar^2}-\frac{1}{2\c^2}
 &-&3{\pi}G\gamma^2\left[\Big(v^2+\frac{{d}^2}{2}\Big)
 \Big(\sin^2{(2b)}\big(1-\frac{4}{d^2}\big)+\frac{2}{{d}^2}\Big)
 +iv\sin{(4b)}\big(1-\frac{4}{d^2}\big)\right]\nonumber\\
 &&+\frac{3{\pi}G(1+\gamma^2)}{4}\left[\Big(v^2+\frac{{d}^2}{2}\Big)
 \Big(\sin^2{(4b)}\big(1-\frac{16}{d^2}\big)+\frac{8}{{d}^2}\Big)
 +2iv\sin{(8b)}\big(1-\frac{16}{d^2}\big)\right].\label{eqn:effective constraint LQC}
 \ea
Note that $\frac{d^2}{2}$ and $\frac{2}{d^2}$ are the square of
fluctuations of $\hat{v}$ and $\widehat{\sin{(2b)}}$ respectively.
They can be seen as quantum corrections to the leading term:
$v^2\sin^2{(2b)}+iv\sin{(4b)}$ of the \emph{Euclidean} part $\TE$ as well as
the leading term: $v^2\sin^2{(4b)}+2iv\sin{(8b)}$ of the \emph{Lorentz} part.

A careful observation reveals that the real and imaginary parts of the leading
terms can be synthesized into a \emph{Moyal} $*$-\emph{product}
\cite{book-fuzzy}, i.e.,
\begin{subequations}
\ba
 &&v^2\sin^2(2b)+iv\sin{(4b)}=ve^{\frac{i}{2}\left(\overleftarrow{\partial_v}\overrightarrow{\partial}_b
 -\overleftarrow{\partial_b}\overrightarrow{\partial}_v\right)}
 \big(\sin{(2b)v}\sin{(2b)}\big)
=:v*\big(\sin{(2b)v}\sin{(2b)}\big)\label{eqn:LQCE*prod},\\
&&v^2\sin^2(4b)+2iv\sin{(8b)}
 =v*\big(\sin{(4b)v}\sin{(4b)}\big)\label{eqn:LQC*prod}.
 \ea
  \end{subequations}
Therefore the effective Hamiltonian constraint takes the form:
  \ba
 &&{{\mathscr{H}}}_{\rm{eff}}=-\frac{p_{\phi}^2}{\hbar^2}-\frac{1}{2\c^2}
 -3{\pi}G\gamma^2\left[v*\Big(\sin{(2b)v}\sin{(2b)}\Big)\big(1-\frac{4}{d^2}\big)
 +\frac{\sin^2{(2b)}d^2}{2}\big(1-\frac{4}{d^2}\big)+\frac{2v^2}{d^2}+1\right]\nonumber\\
 &&\quad\quad\quad\quad\quad\quad\quad
 +\frac{3{\pi}G(1+\gamma^2)}{4}\left[v*\Big(\sin{(4b)v}\sin{(4b)}\Big)\big(1-\frac{16}{d^2}\big)
 +\frac{\sin^2{(4b)}d^2}{2}\big(1-\frac{16}{d^2}\big)+\frac{8v^2}{d^2}+4\right].
 \label{eqn:effective constraint LQC with *}
 \ea
The \emph{Moyal} $*$-\emph{product} emerges in the gravitational
part of the Hamiltonian, since both
$\TE\propto\hat{v}(\widehat{\sin{(2b)}}\hat{v}\widehat{\sin{(2b)}})$
and
$\TL\propto\hat{v}(\widehat{\sin{(4b)}}\hat{v}\widehat{\sin{(4b)}})$
are non-symmetric operators which can be regarded as a product of
two self-adjoint operators. Thus, in this model the coherent state functional
integral also suggests the \emph{Moyal} $*$-\emph{product} to express the
effective Hamiltonian for the quantum system with a
\emph{non-symmetric} Hamiltonian operator. However,
since the \emph{Moyal} $*$-\emph{product} originates from the non-symmetry of the operator, one may doubt why we did not use a \emph{symmetric} operator from the very beginning. To understand the motivation of the non-symmetric operator $\hat{\Theta}$, we recall that the initial Hamiltonian constraint operator in LQC is actually self-adjoint in the kinematical Hilbert space\cite{aps1}. To resolve the constraint equation and find \emph{physical} states, one feasible method is to rebuild the constraint equation as a \emph{Klein-Gordon} like equation and treat the scalar $\phi$ as an \emph{internal time}. As a result, the constrained quantum system was recast into a relativistic particle whose dynamics is govern by a \emph{Klein-Gordon} like equation with an \emph{emergent time} variable \cite{aps2}. The price to get this Klein-Gordon like equation is that the new gravitational Hamiltonian operator $\hat{\Theta}$ becomes a multiplication of two self-adjoint operators, and hence it is no longer symmetric. But this does not indicate that one could not employ $\hat{\Theta}$ in the intermediate step to find physical states. On the other hand, because the \emph{Moyal} $*$-\emph{product} comes from the expectation value of the \emph{multiplication} of two self-adjoint operators on coherent state, this non-symmetric $\hat{\Theta}$ just provides a suitable arena to examine the \emph{Moyal} $*$-\emph{product} from the path integral perspective.

We can also take another practical way to symmetrize $\hat{\Theta}$ at the beginning. For example, one can define a symmetric version of $\TE$ by
\ba
\TE^{'}:=\frac{1}{2}(\TE+\TE^{\dag})
\propto[\hat{v}(\widehat{\sin{(2b)}}\hat{v}\widehat{\sin{(2b)}})+(\widehat{\sin{(2b)}}\hat{v}\widehat{\sin{(2b)}})\hat{v}],
 \ea
 and then carry out the same procedure of above coherent state functional integral. In the calculation of matrix element $\opelem{\cog{n}}{\TE^{'}}{\cog{n-1}}$, we could think that the operators $\hat{v}$ and $\widehat{\sin{(2b)}}\hat{v}\widehat{\sin{(2b)}}$ in $\TE$ act on \emph{bra} $\bra{\cog{n}}$ and \emph{ket} $\ket{\cog{n-1}}$ respectively, while $\widehat{\sin{(2b)}}\hat{v}\widehat{\sin{(2b)}}$ and $\hat{v}$ in $\TE^{\dag}$ act on \emph{bra} $\bra{\cog{n}}$ and \emph{ket} $\ket{\cog{n-1}}$ respectively. Then it is not difficult to see that the imaginary parts generated by $\TE$ and $\TE^{\dag}$ cancel each other. Hence for the symmetric Hamiltonian operator corresponding to $\hat{\Theta}$, we can finally get the following effective Hamiltonian constraint:
 \be\label{eqn:effective constraint LQC remove *}
 \mathcal{H}:=-\frac{p^2_{\phi}}{\hbar^2}-\frac{1}{2\c^2}+3\pi{G}\left(v^2+\frac{1}{2\varepsilon^2}\right)
 \Big(\sin^2{(2b)}\big(1-(16+12\gamma^2)\varepsilon^2
 -(1+\gamma^2)(1-16\varepsilon^2)\sin^2{(2b)}\big)+2\varepsilon^2\Big),
 \ee
which takes the same form as (\ref{eqn:effective constraint LQC with *}) but without $*$-\emph{product}  while ${\varepsilon}\equiv1/d$ denotes the quantum fluctuation of
$\sin b$.

\section{Effective dynamics}
Using the effective Hamiltonian constraint $\mathscr{H}_{\rm{eff}}$ which contains \emph{Moyal} $*$-\emph{product},
one may investigate the corresponding dynamics by defining the evolution equation as:
 \be\label{eqn:evolution with *}
 \dot{f}(v,b):=\frac{1}{i\hbar}\left(f*{\mathscr{H}}_{\rm{eff}}-{\mathscr{H}}_{\rm{eff}}*f\right),
 \ee
for any dynamical quantity $f(v,b)$. Especially, the evolution of basic variables can be obtained as:
 \begin{subequations}
 \ba
 \dot{v}&=&-\frac{12\pi{G}\gamma^2}{\hbar}\Big[v*\big(v\sin{(2b)}\cos{(2b)}(1-4{\varepsilon}^2)\big)
 +\frac{\sin{(2b)}\cos{(2b)}(1-4{\varepsilon}^2)}{2{\varepsilon}^2}\nonumber\\
 &&\quad\quad\quad\quad\quad\quad+\left(\frac{v^2}{2}-\Big(v^2+\frac{1}{2{\varepsilon}^2}\Big)\sin^2{(2b)}
 -\frac{\sin^2{(2b)}(1-4{\varepsilon}^2)}{8{\varepsilon}^4}\right)
 \partial_b{\varepsilon}^2\Big]\nonumber\\
 &&+\frac{12\pi{G}(1+\gamma^2)}{4\hbar}\Big[2v*\big(v\sin{(4b)}\cos{(4b)}(1-16{\varepsilon}^2)\big)
 +\frac{\sin{(4b)}\cos{(4b)}(1-16{\varepsilon}^2)}{{\varepsilon}^2}\nonumber\\
 &&\quad\quad\quad\quad\quad\quad\quad\quad+\left(2v^2-\Big(v^2+\frac{1}{2{\varepsilon}^2}\Big)4\sin^2{(4b)}
 -\frac{\sin^2{(4b)}(1-16{\varepsilon}^2)}{8{\varepsilon}^4}\right)
 \partial_b{\varepsilon}^2\Big],\label{eqn:v evolution with *}\\
 \dot{b}&=&\frac{3\pi{G}}{\hbar}\Big[\Big(2v\big(1-4{\varepsilon}^2\big)\sin{(2b)}\Big)*\sin{(2b)}
 +{4v}{\varepsilon}^2\nonumber\\
 &&\quad\quad\quad\quad\quad-\Big(\frac{\sin^2{(2b)}}{2\varepsilon^4}\big(1-{4}{\varepsilon}^2\big)
 +\big(v^2+\frac{1}{2{\varepsilon}^2}\big){4\sin^2{(2b)}}-{2v^2}\Big)\partial_v{\varepsilon}^2\Big]\nonumber\\
 &&-\frac{3\pi{G}(1+\gamma^2)}{4\hbar}\Big[\Big(2v\big(1-16{\varepsilon}^2\big)\sin{(4b)}\Big)*\sin{(4b)}
 +{16v}{\varepsilon}^2\nonumber\\
 &&\quad\quad\quad\quad\quad\quad\quad\quad-\Big(\frac{\sin^2{(4b)}}{2\varepsilon^4}\big(1-{16}{\varepsilon}^2\big)
 +\big(v^2+\frac{1}{2{\varepsilon}^2}\big){16\sin^2{(2b)}}-{8v^2}\Big)\partial_v{\varepsilon}^2\Big]
 \label{eqn:b evolution with *}
 \ea
 \end{subequations}
where $\partial_b{\varepsilon}^2\equiv\partial({\varepsilon}^2)/\partial b$ and $\partial_v\equiv \partial/\partial v$.
However, there seems no way to understand Eqs. (\ref{eqn:v evolution with *}) and (\ref{eqn:b evolution with *}) directly as effective classical equations because of the $*$-\emph{product} therein. Instead, we could use the effective Hamiltonian constraint (\ref{eqn:effective constraint LQC remove *}) without $*$-\emph{product} to explore the effective dynamics. Using the conventional Poisson bracket, we can derive a modified Friedmann equation from the effective Hamiltonian (\ref{eqn:effective constraint LQC remove *}) as:
 \ba
 &&H^2\equiv\left(\frac{\dot{a}}{a}\right)^2
 =\frac{8\pi{G}\rho_{\rm{c}}}{3}\Big[\left(1+\frac{1}{2v^2{\varepsilon}^2}\right)\Big(\sin{(2b)}\cos{(2b)}
  \big(1-(16+12\gamma^2)\varepsilon^2-2(1+\gamma^2)(1-16\varepsilon^2)\sin^2{(2b)}\big)\nonumber\\
 &&\quad\quad\quad\quad\quad\quad\quad\quad\quad\quad\quad\quad\quad\quad\quad\quad\quad\quad\quad
 -\Big(\sin^2{(2b)}\big(4(1+\gamma^2)\cos^2{(2b)}-\gamma^2\big)-\frac{1}{2}\Big)\partial_b\varepsilon^2\Big)\nonumber\\
  &&\quad\quad\quad\quad\quad\quad\quad\quad\quad\quad\quad\quad
  -\Big(\sin^2{2b}\big(1-(16+12\gamma^2)\varepsilon^2-(1+\gamma^2)(1-16\varepsilon^2)\sin^2{(2b)}\big)+2\varepsilon^2\Big)
  \frac{\partial_b\varepsilon^2}{8\varepsilon^2v^2\varepsilon^2}\Big]^2\label{eqn:Fdm eq}
  \ea
where $\rho_c\equiv\frac{\sqrt3}{32\pi^2G^2\hbar\gamma^3}$ is a
constant. To annihilate $\sin{(2b)}$ and $\cos{(2b)}$ in \eqnref{eqn:Fdm eq},
we use the constraint equation (\ref{eqn:effective constraint LQC remove *}) to get
 \be\label{eqn:sin and rho}
 \sin^2{(2b)}=\frac{1-(16+12\gamma^2)\varepsilon^2-\sqrt{\big(1-(16+12\gamma^2)\varepsilon^2\big)^2
 -4(1+\gamma^2)(1-16\varepsilon^2)\chi}}
 {2(1+\gamma^2)(1-16\varepsilon)^2},
 \ee
and
\be\label{eqn:chi}
 \chi\equiv\frac{K}{J}\frac{\rho}{\rho_{\rm{c}}}-2\varepsilon^2,
\ee
where $\rho=\frac{p^2_{\phi}}{2V^2}$ is the density of matter, $K\equiv1+\frac{\hbar^2}{2\c^2p^2_{\phi}}$, and $J\equiv1+\frac{1}{2v^2{\varepsilon}^2}$.
However, \eqnref{eqn:Fdm eq} looks problematic since it depends on the volume $v$ of the
chosen fiducial cell. This originates from the fact that we have to
use the coherent states peaked on the phase points $(v, b)$ in the
path integral. In the final picture we have to remove the infrared
regulator by letting the cell occupy full spatial manifold. In this
limit, the irrelevant correction terms proportional to
$1/(v{\varepsilon})^2$ could be neglected, while the relevant terms
proportional to ${\varepsilon}^2$ would be kept, since ${\varepsilon}$
was understood as the fluctuation of $\sin b$ which does not depend
on the fiducial cell. We finally get
\ba
&&H^2=\frac{8\pi{G}\rho_{\rm{c}}}{3}\Big[\pm\frac{1}{2\beta}\sqrt{\left(\lambda-\sqrt{\lambda^2-4\beta\chi}\right)
\left(2\beta-\lambda+\sqrt{\lambda^2-4\beta\chi}\right)\Big(\lambda^2-4\beta\chi\Big)}\nonumber\\
 &&\quad\quad\quad\quad\quad\quad\quad\quad\quad\quad
 +\Big(\frac{1}{2}-\frac{\lambda-\sqrt{\lambda^2-4\beta\chi}}{2\beta}
 \big(\frac{4(1+\gamma^2)(2\beta-\lambda+\sqrt{\lambda^2-4\beta\chi})}{2\beta}-\gamma^2\big)
 \Big)\partial_b\varepsilon^2\Big]^2\label{eqn:Fdm eq2}
 \ea
where the positive and negative signs correspond to the expanding and contracting universe respectively. Here we use notations: $\lambda\equiv1-(16+12\gamma^2)\varepsilon^2$ and $\beta\equiv(1+\gamma^2)(1-16\varepsilon^2)$ for a concise writing.
Note that \eqnref{eqn:Fdm eq2} implies significant departure from classical GR. For simplicity, we first consider the case $\partial_b\varepsilon^2=0$ and see whether the \emph{bounce} or \emph{re-collapse} determined by $H=0$ could occur. Then it is obvious that, for a contracting universe, the so-called quantum bounce of LQC will occur when
 \be\label{eqn:solution chi 1}
 \chi=\frac{\lambda^2}{4\beta},
 \ee
which means that $\rho$ increases to $\rho_{\rm{boun}}\approx\frac{\rho_{\rm{c}}}{4(1+\gamma^2)}$
if $\varepsilon^2$ is neglected. On the other hand, for an expanding universe, a re-collapse would occer when $\chi=0$, or equivalently $\rho$ decreases to $\rho_{\rm{coll}}\approx2\varepsilon^2\rho_{\rm{c}}$, which coincide with the result in canonical theory \cite{Yang}, where $\partial_b\varepsilon^2$ was assumed as higher order term and hence neglected. As pointed out in Refs.\cite{Yang,Ding}, the inferred re-collapse is almost in all probability as viewed from the parameter space characterizing the quantum fluctuation $\varepsilon$. Intuitively, as the universe expands unboundedly, the matter density would become so tiny that its effect could be comparable to that of quantum fluctuations of the space-time geometry. Then the Hamiltonian constraint may force the universe to contract back. It should be noted that the effective equation and hence the inferred effect of re-collapse are only valid with the coherent states of Gaussian type. Whether there is a similar result for other semiclassical states is still an interesting open issue. For example, one may consider the affine coherent states in the \emph{affine quantum gravity} approach developed by Klauder \cite{Klauder1,Klauder4}.

In the case when $\partial_b\varepsilon^2$ could not be neglected, the bounce would also be approached for a contracting universe. Because $\partial_b\varepsilon^2$ would not be bigger than the order of $\mathcal{O}(\varepsilon^2)$, $\chi$ could be infinitely close to the result in \eqnref{eqn:solution chi 1} and lead to $H=0$. However, for an expanding universe, both of the two terms in the bracket of the right hand side of \eqnref{eqn:Fdm eq2} are non-negative in large scale. As a result, the Hubble parameter would always keep non-zero unless $\partial_b{\varepsilon}^2$ approaches $0$ asymptotically. Therefore, the inferred re-collapse might occur only if $\partial_b{\varepsilon}^2$ approaches $0$ asymptotically.

If we neglect all the higher-order quantum corrections: $\frac{1}{2\c^2}$, $\varepsilon^2$ and $\partial_b\varepsilon^2$, \eqnref{eqn:chi} would be simplified to $\chi=\frac{\rho}{\rho_{\rm{c}}}$, and hence a \emph{first-order} modified Friedmann equation could be obtained from \eqnref{eqn:Fdm eq2} as:
 \be\label{eqn:Fdm eq3}
 H^2=\frac{8\pi{G}\rho}{3}
 \left[1-\frac{\gamma^2+4(1+\gamma^2)\rho/\rho_{\rm{c}}}{1+\gamma^2}
 +\frac{\gamma^2\rho_{\rm{c}}}{2(1+\gamma^2)^2\rho}\left(1-\frac{4(1+\gamma^2)\rho}{\rho_{\rm{c}}}\right)
 \left(1-\sqrt{1-\frac{4(1+\gamma^2)\rho}{\rho_{\rm{c}}}}\right)\right].
 \ee
Note that this first-order modified Friedmann equation is different from that in Refs.\cite{aps1,Ding}. But it coincides with the modified Friedmann equation in Ref.\cite{Yang}. Hence \eqnref{eqn:Fdm eq3} still contains the particular information of alternative dynamics. It is easy to see that if the matter density increase to $\rho=\frac{\rho_{\rm{c}}}{4(1+\gamma^2)}$, Hubble parameter would be zero and the \emph{bounce} could occur for a contracting universe. On the other hand, in the classical regime of large scale, we have $\chi\ll1$ for $\rho\ll{\rho_{\rm{c}}}$ and hence \eqnref{eqn:Fdm eq3} reduces to the standard classical Friedmann equation: $H^2=8\pi{G}\rho/3$.

\section{Summary}
Since there are quantization ambiguities in constructing the
Hamiltonian constraint operator in LQC, it is crucial to check whether the key features of LQC, such as the quantum bounce and effective scenario, are robust against the ambiguities. Moreover, since LQC serves as a simple
arena to test ideas and constructions induced in the full LQG, it is important to implement those treatments from the full theory to LQC as more as possible. Unlike the usual treatment in spatially flat and homogeneous models, the Lorentz term has to be quantized in a form quite different from the Euclidean one in full LQG. For the above purpose, this
kind of quantization procedure which kept the distinction of the Lorentz and Euclidean terms was proposed as alternative dynamics for LQC \cite{Yang}. It was shown in the resulted canonical effective theory that the classical big bang is again replaced by a quantum bounce and it is possible for the expanding universe to re-collapse due to the
quantum gravity effect by certain assumption. Hence it is desirable to study such kind of predictions from different perspective. Meanwhile, it is also desirable to study the \emph{Moyal} $*$-\emph{product} by coherent state functional integral approach within LQC models.

To carry out the above ideas, the present paper is devoted to study the coherent state functional integral in spatially flat isotropic FRW model coupled with a massless scalar field in the alternative dynamics framework of LQC. The main results can be summarized as follows. By the well-established canonical theory, the coherent state functional integral for LQC with alternative dynamics has been formulated by group averaging. For the non-symmetric gravitational Hamiltonian constraint operator, the \emph{Moyal} $*$-\emph{product} emerges naturally in the resulted effective Hamiltonian with higher-order quantum corrections. For the corresponding symmetrized Hamiltonian operator, the effective Hamiltonian and modified Friedmann equation are also derived from the coherent state functional integral approach. It turns out that the quantum bounce resolution of big bang singularity can also be obtained by the path integral representation. On the other hand, if higher order corrections are included, there is a possibility for the re-collapse of an expanding universe due to the quantum gravity effect, which coincides with the result obtained in the canonical formalism.
Moreover, the first-order modified Friedmann equation still contains the particular information of alternative dynamics and hence admits the possible phenomenological distinction between the different proposals of quantum dynamics. The alternative modified Friedmann equation (\ref{eqn:Fdm eq2}) or (\ref{eqn:Fdm eq3}) sets up a new arena for studying phenomenological issues of LQC.

\section*{Acknowledgement}

We thank Abhay Ashtekar and Peng Xu for
helpful suggestion and discussion. This work is supported by NSFC
(No.10975017) and the Fundamental Research Funds for the Central
Universities.

\appendix

\section{Calculation of the functional integral}

We give some details on the calculation of the Lorentz part of the matrix element of exponentiated gravitational Hamiltonian operator:
$\opelem{\cog{n}}{e^{-i\e\a\TL}}{\cog{n-1}}$ .
The order of $\mathcal{O}(\e)$ of this matrix element is
 \ba
 \opelem{\cog{n}}{-i\e\a\TL}{\cog{n-1}}&=&i\e\a\frac{3\pi{G}(1+\gamma^2)}{16}
 \sum_{k}\Big[k(k+4)\cog{n}^*(k)\cog{n-1}(k+8)
 -2k^2\cog{n}^*(k)\cog{n-1}(k)\nonumber\\
 &&\quad\quad\quad\quad\quad\quad\quad\quad\quad\quad\quad
 +k(k-4)\cog{n}^*(k)\cog{n-1}(k-8)\Big]\nonumber\\
 &\equiv&i\e\a\frac{3\pi{G}(1+\gamma^2)}{16}
 \Big(L^+_{n,n-1}-L^0_{n,n-1}+L^-_{n,n-1}\Big).\label{eqn:app-matirx-element}
 \ea
Now we need to deal with the three terms
$L^+_{n,n-1},L^0_{n,n-1},L^-_{n,n-1}$ separately.
First, we get
 \ba
 L^+_{n,n-1}&\equiv&\sum_{k}(k^2+4k)e^{-\frac{(k-v_n)^2}{2d_n^2}-\frac{(k+8-v_{n-1})^2}{2d^2_{n-1}}}
 e^{-ib_n(k-v_n)+ib_{n-1}(k+8-v_{n-1})}\nonumber\\
 &=&\exp{\left(-\frac{8(v_n-v_{n-1})}{d^2_n+d^2_{n-1}}-\frac{32}{d^2_n+d^2_{n-1}}+i8\overline{b}_n
 -\frac{(v_n-v_{n-1})^2}{2(d^2_n+d^2_{n-1})}+i\overline{b}_n(v_n-v_{n-1})\right)}\nonumber\\
 &&\quad\cdot\sum_k(k^2+4k)\exp{\left(-(k-\overline{v}^{+}_n)^2/\overline{d^2_n}
 -i(b_n-b_{n-1})(k-\overline{v}^+_n)\right)},\label{eqn:D+}
 \ea
where
$\overline{v}^+_n\equiv\frac{d^2_{n-1}v_n+d^2_n(v_{n-1}-8)}{d^2_n+d^2_{n-1}}$.
To do the summation in the above equation, we first have to rewrite
$k^2+4k$ as a function of $k-\overline{v}^+_n$:
 \ba
 k^2+4k&=&(k-\overline{v}^+_n)^2+(2\overline{v}^+_n+4)(k-\overline{v}^+_n)
 +{\overline{v}^+_n}^2+4\overline{v}^+_n\nonumber\\
&=&(k-\overline{v}^+_n)^2
+2\left(\overline{v}_n-\frac{6d^2_n-2d^2_{n-1}}{d^2_n+d^2_{n-1}}\right)(k-\overline{v}^+_n)
+{\overline{v}_n}^2-\overline{v}_n\frac{12d^2_n-4d^2_{n-1}}{d^2_n+d^2_{n-1}}
+\frac{32d^2_n(d^2_n-d^2_{n-1})}{(d^2_n+d^2_{n-1})^2},\nonumber
 \ea
where $\overline{v}_n\equiv\frac{d^2_{n-1}v_n+d^2_nv_{n-1}}{d^2_n+d^2_{n-1}}$ .
Do the summation term by term, we get
 \be
 L^+_{n,n-1}=\inner{\cog{n}}{\cog{n-1}}e^{-\frac{32}{d^2_n+d^2_{n-1}}}e^{i8\overline{b}_n}
 \left((\overline{v}_n)^2-\overline{v}_n\frac{8d^2_n}{d^2_n+d^2_{n-1}}+\frac{\overline{d^2_n}}{2}
 +P^+_{n,n-1}\right),
 \ee
where $P^+_{n,n-1}$ denotes a polynomial of $v_n-v_{n-1}$,
$b_n-b_{n-1}$ and $d_n-d_{n-1}$ without the zeroth order term. Here
we have expanded the factor
$\exp{\left(-\frac{4(v_n-v_{n-1})}{d^2_n+d^2_{n-1}}\right)}$ in
\eqnref{eqn:D+} as
$1-\frac{4(v_n-v_{n-1})}{d^2_n+d^2_{n-1}}+\cdots$. Except for the
leading term $1$, all the other terms can be conflated with
$P^+_{n,n-1}$. Under the continuous
limit $N\longrightarrow\infty$, this $P^+_{n,n-1}$ does not
contribute to the effective action of gravity. It is easy to calculate $L^0_{n,n-1}$
and $L^-_{n,n-1}$ as follows:
 \ba
&&L^0_{n,n-1}=2\inner{\cog{n}}{\cog{n-1}}
 \left((\overline{v}_n)^2+\frac{\overline{d^2_n}}{2}
 +P^0_{n,n-1}\right),\\
 &&L^-_{n,n-1}=\inner{\cog{n}}{\cog{n-1}}e^{-\frac{32}{d^2_n+d^2_{n-1}}}e^{-i8\overline{b}_n}
 \left((\overline{v}_n)^2+\overline{v}_n\frac{8d^2_n}{d^2_n+d^2_{n-1}}+\frac{\overline{d^2_n}}{2}
 +P^-_{n,n-1}\right).
 \ea
Taking the expansion
$e^{\left(-\frac{8}{d^2_n+d^2_{n-1}}\right)}=1-\frac{8}{d^2_n+d^2_{n-1}}+\mathcal{O}\left(\frac{1}{d^4}\right)$
and neglecting the higher order terms than
$\left(\frac{1}{d^2}\right)$, we can get the combination
 \ba
 &&L^+_{n,n-1}-L^0_{n,n-1}+L^-_{n,n-1}\nonumber\\
 &=&-4\inner{\cog{n}}{\cog{n-1}}\Big[
\Big((\overline{v}_n)^2+\frac{\overline{d^2_n}}{2}\Big)
\Big(\sin^2{(4\overline{b}_n)}\big(1-\frac{32}{d^2_n+d^2_{n-1}}\big)+\frac{16}{d^2_n+d^2_{n-1}}\Big)\nonumber\\
 &&\quad\quad\quad\quad\quad\quad\quad\quad
 +2i\sin{(8\overline{b}_n)}\overline{v}_n\big(1-\frac{32}{d^2_n+d^2_{n-1}})\frac{2d^2_n}{d^2_n+d^2_{n-1}}\big)
 +P^{\textrm{grav}}_{n,n-1}\Big],\nonumber
 \ea
and hence the matrix element
$\opelem{\cog{n}}{e^{-i\e\a\TL}}{\cog{n-1}}$ is
 \ba
\inner{\cog{n}}{\cog{n-1}}\exp\Big[-i\e\a_n\frac{3\pi{G}(1+\gamma^2)}{4}\Big(
\big((\overline{v}_n)^2+\frac{\overline{d^2_n}}{2}\big)
\big(\sin^2{(4\overline{b}_n)}\big(1-\frac{32}{d^2_n+d^2_{n-1}}\big)+\frac{16}{d^2_n+d^2_{n-1}}\big)&&\nonumber\\
  +2i\sin{(8\overline{b}_n)}\overline{v}_n\big(1-\frac{32}{d^2_n+d^2_{n-1}}\big)\frac{2d^2_n}{d^2_n+d^2_{n-1}}
 +P^{\textrm{L}}_{n,n-1}\Big)\Big].&&\label{eqn:app-L-element}
 \ea
Similarly, we can get the Euclidean part $\opelem{\cog{n}}{e^{-i\e\a{\TE}}}{\cog{n-1}}$ as:
 \ba
\opelem{\cog{n}}{e^{-i\e\a\TE}}{\cog{n-1}}=\inner{\cog{n}}{\cog{n-1}}\exp\Big[i\e\a_n{3\pi{G}\gamma^2}\Big(
\big((\overline{v}_n)^2+\frac{\overline{d^2_n}}{2}\big)
\big(\sin^2{(2\overline{b}_n)}\big(1-\frac{8}{d^2_n+d^2_{n-1}}\big)+\frac{4}{d^2_n+d^2_{n-1}}\big)&&\nonumber\\
  +i\sin{(4\overline{b}_n)}\overline{v}_n\big(1-\frac{8}{d^2_n+d^2_{n-1}}\big)\frac{2d^2_n}{d^2_n+d^2_{n-1}}
 +P^{\textrm{E}}_{n,n-1}\Big)\Big].&&\label{eqn:app-E-element}
\ea
Combining Eqs. (\ref{eqn:app-L-element}) and (\ref{eqn:app-E-element}), we can get \eqnref{eqn:prod LQC result}.

\end{document}